\documentclass[aps,twocolumn,showpacs,superscriptaddress,eps2pdf,reprint]{revtex4-1}

\usepackage{psfrag}
\usepackage{epsfig}
\usepackage{amsfonts}
\usepackage{amssymb}
\usepackage{mathrsfs}
\usepackage{theorem}
\usepackage{amsmath}
\usepackage{times}
\usepackage{color}
\usepackage{soul}
\usepackage{stmaryrd}

\definecolor{jens}{rgb}{.2,0.7,.9}
\definecolor{earl}{rgb}{.7,0.1,.9}

\usepackage{ifpdf}
\ifpdf
\usepackage{epstopdf}   
\fi

\newcommand{\ket}[1]{\ensuremath{\vert#1\rangle}}
\newcommand{\bra}[1]{\ensuremath{\langle #1\vert}}

\renewcommand{\deg}[0]{\ensuremath{\mathrm{deg}}}

\newcommand{\QRM}{\ensuremath{\mathcal{QRM}}}

\def\id{\mbox{\small 1} \!\! \mbox{1}}

\def\id{{\mathchoice {\rm 1\mskip-4mu l} {\rm 1\mskip-4mu l} {\rm 1\mskip-4.5mu l} {\rm 1\mskip-5mu l}}}

\begin{document}
\title{Enhanced fault-tolerant quantum computing in $d$-level systems}

\author{Earl T.\ Campbell}
\affiliation{Department of Physics \& Astronomy, University of Sheffield, Sheffield, S3 7RH, United Kingdom.}
\affiliation{Dahlem Center for Complex Quantum Systems, Freie Universit\"{a}t Berlin, 14195 Berlin, Germany.}


\begin{abstract}

Error correcting codes protect quantum information and form the basis of fault-tolerant quantum computing. Leading proposals for fault-tolerant quantum computation require codes with an exceedingly rare property, a transversal non-Clifford gate. Codes with the desired property are presented for $d$-level, qudit, systems with prime $d$.  The codes use $n=d-1$ qudits and can detect up to ${\sim}d/3$ errors. We quantify the performance of these codes for one approach to quantum computation known as magic-state distillation.  Unlike prior work, we find performance is always enhanced by increasing $d$.
\end{abstract}

\maketitle

Quantum error-correction stores information in a subspace of a larger physical Hilbert space and is an efficient method of protecting quantum information from noise.  Repeated measurements and error-corrections keep the information from drifting too far out of the error-correction subspace, also called the codespace.  For robust quantum computation, we must be able to perform gates without leaving a protected codespace or amplifying existing errors.  Fault-tolerance is straightforward for a limited set of gates, the so-called \textit{transversal} gates of the code. Unfortunately, severe constraints exist~\cite{Xie08,Zeng11,Eastin09,Bravyi13} that mean such direct approaches cannot provide gates sufficient for universal quantum computation.  Rather, we must rely on additional techniques to implement further fault-tolerant gates.  

One route to universality is to prepare high-fidelity resource states and then use state-injection to convert the resource state into a fault-tolerant gate~\cite{Kit03,Rauss07,Knill05,BraKit05,Rei01a,Rei02a,Rei03a}.  Reduction of noise in these resource states, sometimes known as magic states~\cite{BraKit05}, requires extensive distillation methods demanding that the  majority of a quantum computer is a dedicated magic state factory~\cite{Fowler12,Jones12}.  Due to the significant resource overhead, maximizing efficiency of these protocols is of paramount importance, and recently many improvements have been made~\cite{Bravyi12,Meier13,Jones13}.  One could try to circumvent this overhead by exploring one of many other ways to achieve universality~\cite{shor96,Knill96,Raussendorf06,Bombin13,Paetznick13,Anderson14,OConnor14}. However, all these proposals have to sacrifice some error correcting capabilities, and so are only viable when physical operations are much less noisy (such as was explicitly shown in Ref.~\cite{Raussendorf06}).  Except for Shor's method~\cite{shor96}, these alternative routes require codes with a rare property, and such codes also play a fundamental role in most magic state distillation (herein MSD) protocols.  Specifically, these codes have as a transversal gate the $U_{\pi/8}$ phase gate, which is special as it is outside the Clifford group yet still closely related to it.

\begin{figure}[b]
    \centering
    \includegraphics{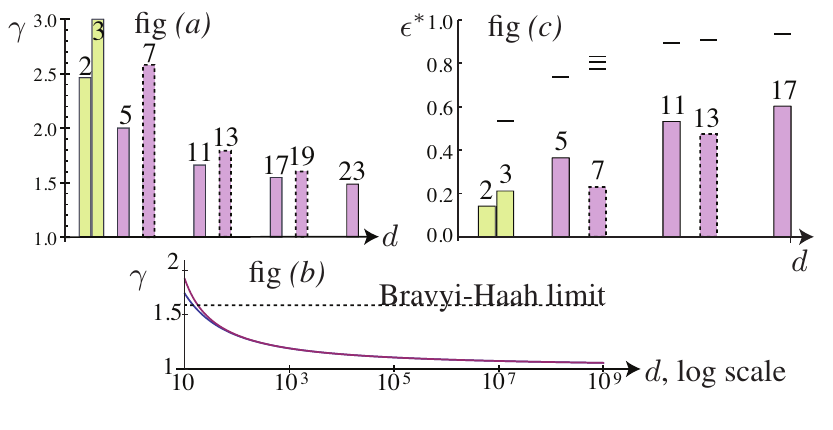} 
    \caption{The performance of our $\QRM$ codes at MSD. \textit{(a)}  shows the efficiency metric $\gamma$ (the smaller the better) for various prime $d$. \textit{(b)} Two curves which $\gamma$ fluctuates between, shown for much larger $d$ (on a log-scale).  \textit{(c)} The depolarizing noise threshold below which the states are distillable. Raised bars show the theoretical maxima where distillation becomes impossible~\cite{Howard11,Veitch12}.  Where multiple maxima appear (see $d=7$), not all magic states $\ket{M_{\mu}}$ are equivalent (see Ref.~\cite{Blanchfield14b} and App.~\ref{App_Cliff_equiv}).}
        \label{Fig_Gamma}
\end{figure}

In almost every route to fault-tolerance, these exotic codes emerge as pivotal components.  Here, we tackle the problem of designing and improving analogous codes in the qudit setting of using $d$-level elementary systems.  In this setting, qudit error-correction~\cite{G02a,G01a,Aharonov} has been long known, but only much later were analogs of $U_{\pi/8}$ phase gate characterized~\cite{Howard12} and codes discovered with these as transversal gates~\cite{Campbell12}.  Campbell, Anwar, and Browne (herein CAB) analyzed several quantifiers of performance for magic state distillation using these codes~\cite{Anwar12,Campbell12}.  CAB found marked improvement over comparable qubit codes for modest sizes $d=3,5$.  However, in even larger dimensions ($d>5$) performance again declined. This fall in performance is peculiar, especially in light of other results showing qudit toric codes have thresholds increasing with system dimension~\cite{Anwar13,Duclos13,Wootton14}.  Here we present MSD protocols with commensurate improvements to thresholds and, more importantly, an unexpected improvement to efficiency that is not present in toric codes.   Fault-tolerant Clifford gates~\footnote{Actually the toric code possesses transversal gates that are a proper subgroup of the Clifford group.} can be provided by toric codes, and our protocols extend this to a universal set of fault-tolerant gates; with both components becoming more effective in higher dimensions, they pose a powerful combination.  Such gains will also apply to qudit analogs of the ``magic-state free" fault-tolerance schemes~\cite{Knill96,Raussendorf06,Bombin13,Paetznick13,Anderson14,OConnor14}.

We consider an extended class of quantum Reed-Muller codes that are constructed in terms of polynomial functions, whereas the work of CAB only considered linear functions.  Here we see that higher degree polynomials can be used to construct codes with the desired transversality properties.  Furthermore, the polynomial degree can grow with the qudit dimension $d$, and, in turn, the effectiveness of the code also grows.  We make this statement more precise by considering the code parameters conventionally labeled $\llbracket n, k, D \rrbracket_d$ where $n$ is the number of physical qudits, $k$ is the number of logical qudits encoded and $D$ is the code distance (measuring its error-correction capabilities).  The codes presented here encode a single logical qudit into $n={d{-}1}$ qudits, and our best performing codes have $D=\lfloor(d+1)/3 \rfloor$.  It is desirable that $D$ is larger and $n$ is smaller, but we find both numbers grow with system dimension.  The overall effectiveness is measured by the codes ``gamma value", $\gamma = \log(n) / \log (D)$, which is smaller for more efficient codes.  We find $\gamma$ can be decreased arbitrarily close to unity by increasing $d$ and discuss the operational meaning of $\gamma$ in the context of MSD.  These results are shown in Fig.~(\ref{Fig_Gamma}), which discuss in detail later.  In summary, they show qudit protocols far ahead of the first proposed qubit codes~\cite{BraKit05}, and comparable with modern qubit block codes~\cite{Bravyi12,Meier13,Jones13} without the disadvantages encurred by using block codes.  

Aside from the practical merits, refinements to notation and proof techniques present a clearer picture of why these codes possess their strange properties.  For technical reasons, we consider only prime dimensions of five and above.  Extensions to prime power dimensions are plausible, and new techniques offer hope in arbitrary dimensions~\cite{Vega14}.  Our results also enhance our understanding of qudit magic states as a resource theory.  Study of qudit systems benefit from neater phase space methods~\cite{Gross06} that are absent from qubit systems, and a growing body of work shows a richer and more elegant resource theory~\cite{Howard11,Veitch12,mari12,Veitch14} than for qubits, with interesting connections to quantum contextuality~\cite{Howard14} and to symmetric informationally-complete measurements~\cite{appleby05,Blanchfield14,Blanchfield14b}. 



\textit{Definitions}.- Qudits are $d$-level systems, and we label computational basis states by elements of the set $\mathbb{F}_{d}=\{ \ket{0}, \dots , \ket{{d{-}1}}\}$ and all arithmetic is performed modulo $d$, so $\mathbb{F}_{d}$ forms a Galois field of order $d$.  Let us review the structure of the Clifford group denoted $\mathcal{C}_{d}$ and a normal subgroup called the Pauli group denoted $\mathcal{P}_{d}$, both being fundamental to quantum coding theory.  Those Clifford unitaries that are also diagonal (in the standard basis) have the form $\mathcal{Z}_{\alpha, \beta}:= \omega^{\alpha \hat{n} + \beta \hat{n}^2 }$
where $\alpha, \beta \in \mathbb{F}_{d}$, throughout $\omega = \exp( i 2 \pi / d) $ and we employ the number operator $\hat{n} = \sum_{x}x \vert x \rangle \langle x \vert$.  Generally, the Clifford exponent is quadratic in the number operator, but when linear we find $\mathcal{Z}_{\alpha, 0}$ is also in the Pauli group.  In particular, Pauli $Z$ is $\mathcal{Z}_{1,0}$. There are also Clifford unitaries that permute the computational basis states, such that $\mathcal{X}_{\alpha,\beta} \ket{x}=\ket{ \alpha  + \beta x }$, where again $\alpha, \beta \in \mathbb{F}_{d}$.  Here, unit $\beta$ picks out elements of the Pauli group, in particular, Pauli $X$ is $\mathcal{X}_{1,1}$.  The Pauli operators $X$ and $Z$ and tensor products thereof generate the whole Pauli group. Multiplication in modular arithmetic is not always invertible, which we need for $\mathcal{X}_{\alpha,\beta}$ to be unitary, but thankfully in prime dimensions we do have invertibility. These gates are not yet sufficient to generate the Clifford group and we must also include a Hadamard-like gate, $H$, that acts as $ H \ket{x}= \sum_{y} \omega^{xy} \ket{y} /\sqrt{d}$ and a 2-qudit control-phase gate of the form $C_{Z}=\omega^{\hat{n} \otimes \hat{n}}$.  

\textit{Non-Clifford gates}.- As remarked earlier, the special ingredient we need is a fault-tolerant implementation of a gate outside of the Clifford group.  The qubit $U_{\pi/8}$ gate is non-Clifford and has other useful properties.  We consider qudit analogs~\cite{Howard12,Campbell12}. Such an analog will be diagonal in the computational basis, non-Clifford, and will by conjugation map Pauli operators to Clifford operators.  Such gates are often said to belong in the third level of the Clifford hierarchy, and this property is useful for gate teleportation~\cite{Gottesman99}.  We show these properties for unitaries of the form, $M_{\mu}:=\omega^{\mu \hat{n}^{3}}$ for $\mu=1,\ldots,d-1$.  We could include a quadratic component to the exponent, but the unitary would be Clifford equivalent (for more insights on Clifford equivalence, see Ref.~\cite{Blanchfield14b} and the App.~\ref{App_Cliff_equiv}).   We find that
 \begin{eqnarray}
\label{M_conjugation}
    M_{\mu} X M_{\mu}^{\dagger} &=& X  \omega^{\mu  ( (\hat{n}+1)^3 - \hat{n}^3  )} = X \omega^{\mu(3 \hat{n}^2 + 3 \hat{n} + 1 )} ,
 \end{eqnarray}
is Clifford and non-Pauli.  It follows that $M_{\mu}$ is in the third level of the Clifford hierarchy, and so analogous to a $U_{\pi/8}$ gate (more details in App.~\ref{App_Hierarchy}).  However, there are important differences between qubits and higher $d$ qudits. For instance, the qudit non-Clifford gates are order $d$ (so $M_{\mu}^{d} = \id$), but the qubit $U_{\pi/8}$ is order $8=2^3$, which changes the proof of transversality and ultimately leads to our improved performance.


\textit{The Reed-Muller codes}.-Here we consider a simple subclass of shortened quantum Reed-Muller codes (herein $\QRM$ codes).  The codes are defined in terms of polynomial functions from the nonzero elements of the field ($\mathbb{F}_{d}^{*}=\{1 , \dots , {d{-}1} \}$) to the whole field, so formally $F:\mathbb{F}^{*}_{d} \rightarrow \mathbb{F}_{d} $.  It is called a degree-$r$ polynomial if
 \begin{equation}
    F(x) = f_{0} + \sum_{m=1,\ldots ,r} f_{m} x^{m} ,
 \end{equation}
with $f_{r}\neq 0$ and where lower case is used throughout for the coefficients $f_{k} \in \mathbb{F}_{d}$.  We also denote the degree as $\mathrm{deg}(F)$.  Fermat's little theorem (FLT) asserts that $x^n=x^m \pmod{d}$ if $n=m \pmod{{d{-}1}}$, and so all functions can be represented with polynomials of degree less than ${d{-}1}$. When higher degree polynomials appear, we equate them via FLT with the polynomial of lowest possible degree. We say the function is shifted by $f_{0}$ and unshifted when $f_{0} = 0$.  We use these functions to describe quantum states, so that
 \begin{equation}
 \label{Ket_func}
     \ket{\psi_{F}} = \ket{ F(1) }\ket{ F(2) } \ldots  \ket{ F(d-2) } \ket{ F(d-1) }
\end{equation}
For example, $\ket{\psi_{F(x)=x}}=\ket{1}\ket{2}...\ket{d-1}$. The codespace of such $\QRM$ codes is defined by its degree, $r$.  We begin by defining the logical states $\ket{k_{L}}$ as
 \begin{equation}
 \label{Eq_kL_expansion}
    \ket{k_{L}} = d^{-r/2} \sum_{\substack{\mathrm{deg}(F) \leq r \\ f_{0} = k}}   \ket{\psi_{F}}  ,
 \end{equation}
which is an equally weighted sum over all $\ket{\psi_{F}}$, where $F$ is a $k$-shifted function of degree no greater than $r$.  For the $\ket{0_{L}}$ state, the functions are unshifted.

Alternatively, $\QRM$ codes can be described in the qudit stabilizer formalism~\cite{Gott99,Beaudrap}.  If a Pauli operator $s$ satisfies $s \ket{k_{L}} = \ket{k_{L}}$ for all $k$, we say that $s$ is an element of the code stabiliser $\mathcal{S}$.  We proceed by defining Pauli operators
 \begin{equation}
    X_{F}=X^{F(1)} \otimes X^{F(2)} \otimes \ldots  \otimes  X^{F(d-1)}  ,
 \end{equation}
and noting $X_{F}\ket{\psi_{G}}=\ket{\psi_{F+G}}$.  One can then verify that the codespace is stabilized by $X_{F}$ for all unshifted functions of degree no greater than $r$.  Similarly, the stabilizer contains operators $Z_{F}$, as we find later when discussing code distance.  In terms of logical operators, we observe that since $X^{\otimes ({d{-}1})}\ket{\psi_{F}}=\ket{\psi_{F+1}}$, we have $X^{\otimes ({d{-}1})}\ket{k_{L}}=\ket{(k+1)_{L}}$ and can identify $\overline{X}=X^{\otimes ({d{-}1})}$ where bars throughout denote logical operators.

\textit{Transversal gates}.-The key feature of $\QRM$ codes that makes them useful for fault-tolerant quantum computing is that they possess transversal non-Clifford gates.  That is, one can perform a logical non-Clifford single qudit gate within the codespace by applying a product unitary. We shall show that for $\QRM$ codes of degree $3r < {d{-}1}$, a logical $M_{\mu}$ gate can be implemented transversely by $\overline{M}_{\mu}= M_{-\mu}^{\otimes ({d{-}1})}$.  We begin by considering some polynomial $F$, the corresponding state $\ket{\psi_{F}}$, and how the product unitary acts on this
 \begin{equation}
 \label{Eq_Mmu}
    \overline{M}_{\mu}\ket{\psi_{F}} =      \omega^{- \mu \sum_{x=1}^{{d{-}1}} F(x)^3 } \ket{\psi_{F}} = \omega^{- \mu S(H)} \ket{\psi_{F}} ,
 \end{equation}
where we introduce the shorthand $S(H):=\sum_{x=1}^{{d{-}1}}H(x)$ and $H(x):=F(x)^3$, and next we must evaluate $S(H)$.  The following steps all rest on a remarkable algebraic feature of prime numbers, namely that all functions satisfy $S(H)=-h_{0}$ (see App.~\ref{APP_SUM}).  Now we must find the explicit form for $h_{0}$ in terms of the $f_{m}$.  By expanding out $F^3$ and using FLT, we find $h_{0}$ is a sum over every $f_{m_{1}}f_{m_{2}}f_{m_{3}}$ where $m_{1}+m_{2}+m_{3}=0 \pmod{{d{-}1}}$. This is hugely simplified if we restrict to $F$ with degree less than $({d{-}1})/3$ as there is only one contribution so that $h_{0}=f_{0}^3$.  Under this assumption, Eq.~(\ref{Eq_Mmu}) becomes simply $\overline{M}_{\mu}\ket{\psi_{F}} = \omega^{- \mu f_{0}^3}\ket{\psi_{F}}$, and so the phase depends only on the shift of the function.  

In Reed-Muller codes, the logical basis states $\ket{k_{L}}$ are a sum over polynomials shifted by $k$.  For $\QRM$ codes of degree $r<(d-1)/3$, these polynomials have degree of $r$ or less, and so the above proof directly entails $ \overline{M}_{\mu} \ket{k_{L}} =     \omega^{-\mu k^{3}} \ket{k_{L}} $. This shows that the product unitary acts on the $\QRM$ codewords as a logical $M_{\mu}$ gate, and so transversality of a non-Clifford gate has been demonstrated.

\textit{Error correcting properties}.- The previous work of CAB~\cite{Campbell12} used different proof techniques to show the transversality of non-Clifford gates for $\QRM$ codes of only first degree.  Furthermore, those first degree codes could only detect a single error.  However, increasing the degree of $\QRM$ codes opens the possibility of detecting more errors.  Indeed, we will show below that $\QRM$ codes of degree $r$ can detect up to $r$ errors.

We first review some basic concepts.  The weight of an operator $P$ denoted $\mathrm{wt}(P)$ is the number of qudits it acts upon non-trivially.  We are interested in the smallest weight Pauli operator whose effects cannot be detected by measuring stabilizers of the code and that also acts non-trivially on the code.  Formally, $P$ must commute with $\mathcal{S}$, but not be a member of $\mathcal{S}$, and the minimum weight of such $P$ is called the code distance,
 \begin{equation}
    D = \min \{ \mathrm{wt}(P) | [P, \mathcal{S}]=0 , P \notin \mathcal{S} ,P \in \mathcal{P}  \} .
 \end{equation}
We begin by considering only phase errors and again use polynomials to define multi-qudit operators, so $Z_{G}=\otimes_{x}Z^{G(x)}$.  The weight of $Z_{G}$ can be expressed as $\mathrm{wt}(Z_{G})=({d{-}1})-\chi(G)$, where $\chi(G)$ is the number of nonzero arguments for which the function $G$ evaluates to zero.  That is, $\chi (G)$ is the number of distinct nonzero roots of the polynomial.  The roots are limited by the degree of the polynomial such that $\chi (G)\leq \mathrm{deg}(G)$.  Putting this together, we have the degree-weight relation $\mathrm{wt}(Z_{G})\geq({d{-}1})-\mathrm{deg}(G)$.  Next, we show that commutation of $Z_{G}$ with the stabilizer puts an upper bound on $\mathrm{deg}(G)$.

Recall that the stabilizers of $\QRM$ codes include operators $X_{F}$ for all unshifted functions with $\mathrm{deg}(F)\leq r$.  From $ZX=\omega XZ$, we know that $X_{F}$ and $Z_{G}$ commute iff $\sum_{x \in \mathbb{F}_{d}^{*}  } F(x)G(x) = 0 $.  Recall that such sums only vanish when the composite polynomial, here $H'(x):=F(x)G(x)$, is unshifted.  The shift of $H'$ is $h'_{0}$, which by expanding $FG$ is a sum over every $f_{m}g_{n}$ such that $m+n=0 \pmod{{d{-}1}}$. Let us just consider monomials $F(x)=x^{q}$, then we have the simplification $h'_{0}= f_{q}g_{({d{-}1})-q}=g_{({d{-}1})-q}$.   By definition, a degree $r$ polynomial has $g_{r}\neq 0$, and so $h'_{0}\neq 0$  whenever $q=({d{-}1})-\mathrm{deg}(G)$.  Since $q=\deg(F)\leq r$, we conclude that provided $({d{-}1})-r\leq \deg(G)$, we can always find an unshifted $F$ [with $\deg(F) \leq r$] such that $FG$ is shifted. This entails that the corresponding $Z_{G}$ fails to commute with at least one element of the stabilizer, namely $X_{F}$.  Conversely, for a $Z_{G}$ error to commute with the stabilizer, it must have degree less than $({d{-}1})-r$, and from the degree-weight relation this entails $\mathrm{wt}(Z_{G})>r$.  Being true for all $Z_{G}$ entails all undetectable phase errors have weight greater than $r$.  As for $X$ errors, a similar analysis shows even greater protection. We conclude that the code distance satisfies $D \geq r+1$, with equality being  easy to confirm.

\textit{MSD}.- We have shown transversal non-Clifford gates for $\QRM$ codes of up to degree, $r=\lfloor ({d{-}2})/3 \rfloor $, and so a distance $D=\lfloor (d+1)/3 \rfloor$ that grows linearly.  This result implies increasing code performance with increasing dimensionality.  To make this concrete, let us consider the performance when using these codes for MSD.  Each round of distillation consumes (${d{-}1}$) noisy copies of $M_{-\mu}\vert + \rangle$ and, when successful, outputs a purer magic state.  To briefly review, a simple distillation protocol will take noisy magic states, measure the code stabilizers, and postselect on all ``+1" outcomes.  This projects onto the codespace.  Next, we perform $\ket{k_{L}}\rightarrow \ket{k} \ket{0}...\ket{0}$, outputting the first qudit.  The role of transversality of $M_{\mu}$ is the following: it commutes with the code projection allowing us to instead consider the action of the projector onto (independently) noisy $\vert + \rangle$ states and consequently directly detect errors therein.  We expand on this last point. The $Z$  basis measurements will project noisy $\vert + \rangle$ states into the codespace, allowing the $X$ basis measurements to detect phase errors.  Consequently, the state is output with an infidelity $\epsilon'  =O(\epsilon^{D}) $ where $D$ is the code distance, as analyzed in detail in Refs.~\cite{BraKit05,Campbell12}. The above argument gives sufficient conditions for a code being useful for MSD.  Codes without the desired transversality might, via a different mechanism, achieve MSD, but the performance is no longer linked to code distance.  For instance, MSD can be based on the 5-qubit code~\cite{BraKit05} or Steane code~\cite{Rei01a} and, respectively, reduces errors quadratically $\epsilon'=O(\epsilon^2)$ and linearly $\epsilon'=O(\epsilon)$, though both codes have $D=3$.  
Only the $X$ basis measurement detect errors, whereas the $Z$ basis measurements are entirely random.  Prior work~\cite{BraKit05,Campbell12,Bravyi12} shows that all $Z$-basis measurements' outcomes can be accepted providing a success probability approaching unity as $\epsilon \rightarrow 0$.


We now discuss the efficiency and noise threshold of MSD with the high degree $\QRM$ codes.  The average number of consumed noisy states can be shown~\cite{Campbell12} to scale as $C\log^{\gamma} ( \epsilon_{\mathrm{final}} )$ where $\gamma= \log_{D}( {d{-}1})$.  Therefore, $\gamma$ quantifies the efficiency for MSD, with lower $\gamma$ showing better performance.  Using the optimal $D=\lfloor (d+1)/3 \rfloor$, we find $\gamma$ slowly approaches 1 from above.  Previously proposed qudit codes were first degree $\QRM$ codes (with distance $2$) and though CAB observed improved $\gamma$ for $d=5$, this was followed by impoverished $\gamma$ at greater $d$.  Whereas, here $\gamma$ continues to improve with growing $d$.  In the qubit setting, the 15-qubit code achieves $\gamma=2.465$.  However, recent improvements in qubit MSD have used block codes~\cite{Meier13,Bravyi12} that achieve $\gamma \rightarrow 1.585$ (the Bravyi-Haah limit) and multi-level distillation~\cite{Jones13} where numerics indicate $\gamma \rightarrow 1$.  However, multi-level distillation requires very many magic states to be simultaneously prepared, and cannot be used in conjunction with further resource saving methods that store nosier magic states within smaller codes.  One detailed study of multi-level distillation concluded that the benefits are slight compared to the 15-qubit code~\cite{fowler13}. In contrast, higher dimensional systems offer efficient protocols even for preparation of a single magic state, avoiding the complexity of multi-level or block protocols.  Though very large $d$ is needed for our protocols to get $\gamma$ close to unity (see Fig.~\ref{Fig_Gamma}), modest $d$ is sufficient to outperform qubit protocols.

Another important figure of merit for MSD is the noise threshold below which the protocol successfully reduces noise.  The threshold depends on the noise model, and here we consider depolarizing noise such that $M_{\mu}\ket{+}$ is mixed with the identity to give a state $\rho$ with fidelity  $\bra{+} M_{\mu}^{\dagger} \rho M_{\mu}\ket{+}= 1-\epsilon$.  After a distillation round, the output is still depolarized with improved fidelity provided $\epsilon < \epsilon^{*}$.  For prime $d \leq 17$, we have numerically found $\epsilon^{*}$ shown in  Fig.~(\ref{Fig_Gamma}c), and observed increasing improvements with $d$.  There is a monotonic improvement in both threshold and $\gamma$ within the two classes of odd numbers, $d =1 \pmod{3}$ versus $d =2 \pmod{3}$.  Jumps occur because the code distance only increases when $d$ increases by 3 or more.  Compared to qudit toric code thresholds, these distillation thresholds are consistently higher, and we exceed $\epsilon^*_\mathrm{dep}=0.5$ even with $d=11$ while toric codes only approach $0.5$ in the large $d$ limit. Even so, there remains potential room for improvement in MSD thresholds, though even in the qubit case, studying the maximum possible threshold is notoriously difficult~\cite{Camp10a}.

\textit{Conclusions}.- In summary, we have shown that quantum Reed-Muller codes provide effective means of fault-tolerantly implementing gates that are essential to various approaches to quantum computing, with special attention paid to MSD.  Further study is warranted of the benefits for qudit variants of non-MSD approaches to quantum computing~\cite{Knill96,Raussendorf06,Bombin13,Paetznick13,Anderson14,OConnor14}, though the relative merits of these proposals is still poorly understood.  Unlike previous work, improvements in efficiency and thresholds continue to increase with $d$.  Comparable threshold improvements are seen in qudit toric code, but improved efficiency does not occur in the toric code context, and so this is the more surprising result. We must remark that coherent control of high $d$ qudits is challenging, and in physical systems one may also see noise rise with $d$.  Such features depend subtly on the details of the underlying physics.  Whilst many systems may not be well suited to qudit approaches, many atomic systems come equipped with large Hilbert spaces for which control of many levels need not be substantially more difficult than control of just 2 levels.  For instance, experiments in trapped cesium have performed gates between 16 levels at $99 \%$ fidelity~\cite{Smith13}.  

\textit{Acknowledgments}.-  We thank the European Commission (SIQS) for funding and Mark Howard, Niel de Beaudrap, Hussain Anwar, Dan Browne, and Tom Bullock for useful discussions and comments on the manuscript.

\appendix

\section{Clifford hierarchy}
\label{App_Hierarchy}

For completeness, we give extra details here on why $M_{\mu}$ (for nonzero $\mu$) is in the third level of the Clifford hierarchy.  First we examine the equation
\begin{equation}
\label{App_Eq1}
    M_{\mu} X M_{\mu}^{\dagger} = X  \omega^{\mu  ( (\hat{n}+1)^3 - \hat{n}^3  )} = X \omega^{\mu(3 \hat{n}^2 + 3 \hat{n} + 1 )} .
\end{equation}
Expanding the exponent, $t:=\mu  ( (\hat{n}+1)^3 - \hat{n}^3  )$, the cubic terms cancel and we are left with a purely quadratic expression, namely $t=\mu(3 \hat{n}^2+ 3 \hat{n}+ 1)$.  Therefore, the RHS is a Clifford that we denote $C=M_{\mu}XM_{\mu}^{\dagger}$.  We further require that $M_{\mu} P M_{\mu}^{\dagger}$ is Clifford for all Pauli $P$.  All Pauli operators can, up to a phase, be written $P=X^m Z^n$ and so conjugate to $M_{\mu}PM_{\mu}^{
\dagger}=C^m Z^n$ where we have used commutation of $M_{\mu}$ with $Z$.  Since the Clifford gates form a group $C^m Z^n$ is again Clifford.

Next we must check that $M_{\mu}$ is not simply a Clifford unitary.  Clifford unitaries conjugate Pauli operators to Pauli operators.  Therefore non-Cliffordness can be established by finding a single Pauli such that $M_{\mu}PM_{\mu}^\dagger$ is not a Pauli.  We again consider $P=X$, and the RHS of Eq.~(\ref{App_Eq1}).  For $C$ to be nonPauil, the exponent $t=\mu(3 \hat{n}^2+3\hat{n} + 1 )$ must be non-linear.  This occurs provided the coefficient of $\hat{n}^2$ is nonvanishing.  Therefore, we need $3 \mu \neq 0$ modulo $d$.  If $d=3$, we have a problem as $3 \mu$ (mod $3$) vanishes for all $\mu$.  Whereas, for odd prime $d>3$, multiplication by 3 is always non-trivial~\footnote{In finite fields, multiplication is always invertible.  Also, the zero element maps to zero, and so no other element can do so.  Hence, typically $3x\neq 0$ when $x\neq 0$.  This argument does not apply when $d=3$ since multiplication by $3$ is actually multiplication by 0, and so not invertible.}.  This is one of the fundamental reasons why it is simpler to construct codes with transversal non-Cliffords in odd $d>3$.  

\section{Clifford equivalence of $M_{\mu}$ gates}
\label{App_Cliff_equiv}

Earlier we remarked that a $M_{\mu}$ gate will be Clifford equivalent to gates of the form $\omega^{\alpha\hat{n} + \beta \hat{n}^2 + \mu \hat{n}^3}$, simply by combining $M_{\mu}$ with a Clifford gate $\mathcal{Z}_{\alpha, \beta}$.  However, this prompts the question whether all $M_{\mu}$ are Clifford equivalent or whether the above $\QRM$ codes ever have genuinely distinct non-Clifford gates.  This question was addressed by Bengtsson \textit{et. al.} in relation to mutually-unbiased bases (MUB) and symmetric, informally-complete (SIC) measurements, and here we present a brief review. In dimension $d=2 \pmod 3$, all $M_{\mu \neq 0}$ gates are Clifford equivalent. Whereas in dimensions where $d=1 \pmod 3$, we find 3 distinct equivalence classes.

We can move between different $M_{\mu}$ gates by conjugating the unitary with the permutation Cliffords $\mathcal{X}_{\alpha, \beta}$ so that
\begin{eqnarray}
    M' &=& \mathcal{X}_{\alpha, \beta}M_{\mu}\mathcal{X}_{\alpha, \beta}^{\dagger} \\ \nonumber
    &=& \omega^{\mu(\alpha + \beta \hat{n})^3}  \\ \nonumber
    &=& \omega^{\mu(\alpha^3 +3 \alpha^2 \beta \hat{n} + 3 \alpha \beta^2 x^2 + \beta^3 \hat{n}^3)}  \\ \nonumber
    &=& \omega^{\mu \beta^3 \hat{n}^3 + g(\hat{n})} ,
\end{eqnarray}
where in the last line we collect all the quadratic terms into $g( \hat{n} )=\mu(\alpha^3 +3 \alpha^2 \beta \hat{n} + 3 \alpha \beta^2 \hat{n}^2)$.  The interesting part is the cubic term, which has gone from $\mu x^3$ to $\beta^3 \mu x^3$.  Hence, $M_{\mu}$ and $M_{\mu'}$ are Clifford equivalent if there exists a $\beta$ such that $\mu'=\beta^3 \mu$.  The structure of the equivalence class is determined by the set 
\begin{equation}
    R_{d} =   \{ \beta^3 | \beta \in \mathbb{F}_{d} \setminus \{ 0 \}\} ,
\end{equation}
which in field theory is known as the cubic residue of the field.  It can be shown that the cubic residue forms a group under multiplication.  We can now immediately leverage results in field theory.  

In dimensions satisfying $d=2 \pmod 3$, the cubic residue includes all non-zero elements of the finite field $\mathbb{F}_{d}$, so $R_{d} = \mathbb{F}_{d} \setminus \{ 0 \}$.  Therefore, we can always find a $\beta$ such that $\beta^3=\mu /  \mu'$ to ensure Clifford equivalence.  

In dimensions satisfying $d=1 \pmod 3$, elementary field theory shows the cubic residue contains $({d{-}1})/3$ elements.  The cubic residue is a normal subgroup of the group $\mathbb{F}_{d} \setminus \{ 0 \}$ (a group under multiplication).  Therefore, each element of $\mathbb{F}_{d} \setminus \{ 0 \}$ belongs to one coset of the cubic residue, and each coset is equal in size.  This gives three cosets each containing $({d{-}1})/3$ elements.  Cosets are closed under multiplication   by elements from the generating subgroup, the cubic residue, and hence each coset defines an equivalence class of $\mu$ values.  For example, $d=7$ satisfies $d=1 \pmod 3$ and the cubic residue is $R_{7}= \{ 1, 6 \}$ with cosets $\{2,5 \}$ and $\{3,4 \}$, providing the three Clifford equivalence classes of $\mu$ values.  

So far we have only considered changing $M_{\mu}$ by applying Clifford gates that permute and apply phases in the computational basis, having ignored the effect of the Hadamard gate.  However, Hadamard gates simply change the basis, and so Hadamards effectively just interchange the role of subsequent $\mathcal{X}_{\alpha, \beta}$ and $\mathcal{Z}_{\alpha, \beta}$ Clifford unitaries.

\section{Evaluating summations}
\label{APP_SUM}

We wish to find a general solution of 
\begin{equation}
    S(H)=\sum_{x \in \mathbb{F}^*_{d}}H(x),
\end{equation}
for all $H(x)=\sum_{m}h_{m}x^{m}$, and as usual working modulo $d$.  We begin by breaking the sum into monomials, 
\begin{equation}
    S(H) =\sum_{m} h_{m} \left( \sum_{x \in \mathbb{F}^*_{d}}x^m \right) = \sum_{m} h_{m} S(x^m),
\end{equation}
and see the problem reduces to finding the monomial solutions $S(x^m)$.  

First we consider sums for $m \neq 0 \pmod{{d{-}1}}$.  Take a non-zero $y$, such that $y^{m} \neq 1$.  The sum can be reordered to be over $(yx)^m$, so that 
\begin{equation}
    S(x^m) = \sum_{xy \in \mathbb{F}^{*}_{d} } (yx)^m  ,
\end{equation}
we reorder back
\begin{equation}
    S(x^m) = y^{m}\sum_{x \in \mathbb{F}^{*}_{d} } x^m ,
\end{equation}
and so $S(x^m)=y^m S(x^m)$. Since $y^m \neq 1$, we conclude $S(x^m)=0$.  Next we consider $m=0 \pmod{{d{-}1}}$, for which $x^m=1$ for all $x \in \mathbb{F}_{d}^{*}$ (recall the star denotes the absence of zero so we never encounter $0^0$).  The sum has $({d{-}1})$ terms and so obviously $S(x^m)={d{-}1}=-1$.

We have shown that
\begin{equation}
   S(x^m)= \sum_{x=1}^{{d{-}1}} x^m = \bigg\{ \begin{array}{lll}
    0 &\mbox{when } m \neq 0 & \pmod{{d{-}1}} , \\
    -1 &\mbox{when } m=0 & \pmod{{d{-}1}} .
 \end{array}
\end{equation}
The modular dependence on $m$ comes from FLT.  We deduce that if $H$ is unshifted, and so contains no $x^0$ terms, then $S(H)=0$, whereas in general $S(H)=-h_{o}$.

\end{document}